\documentclass[aps,prl,twocolumn,showpacs,superscriptaddress,groupedaddress]{revtex4}  
\usepackage{graphicx}  
\usepackage{dcolumn}   
\usepackage{bm}        
\usepackage{amssymb}   

\usepackage{psfrag}
\usepackage{epsfig}
\usepackage{epsf}
\usepackage{float}
\usepackage{slashed}

\newcommand{\beq}{\begin{equation}}
\newcommand{\eeq}{\end{equation}}
\newcommand{\beqa}{\begin{eqnarray}}
\newcommand{\eeqa}{\end{eqnarray}}
\newcommand{\nn}{\nonumber \\ }

\begin{document}

\title{Precision nucleon-nucleon potential at fifth order in the chiral expansion}

\author{E.~Epelbaum}
\affiliation{Institut f\"ur Theoretische Physik II, Ruhr-Universit\"at Bochum,
  D-44780 Bochum, Germany}
\author{H.~Krebs}
\affiliation{Institut f\"ur Theoretische Physik II, Ruhr-Universit\"at Bochum,
  D-44780 Bochum, Germany}
\author{U.-G.~Mei{\ss}ner}
\affiliation{Helmholtz-Institut~f\"{u}r~Strahlen-~und~Kernphysik~and~Bethe~Center~for~Theoretical~Physics,
~Universit\"{a}t~Bonn,~D-53115~Bonn,~Germany}
\affiliation{Institut~f\"{u}r~Kernphysik,~Institute~for~Advanced~Simulation,
and J\"{u}lich~Center~for~Hadron~Physics, Forschungszentrum~J\"{u}lich,~D-52425~J\"{u}lich,~Germany}
\affiliation{JARA~-~High~Performance~Computing,~Forschungszentrum~J\"{u}lich,~D-52425~J\"{u}lich,~Germany}
\date{\today}

\begin{abstract}
We present a nucleon-nucleon potential at fifth order in chiral effective field
theory.  We find a substantial improvement in the description of 
nucleon-nucleon phase shifts as compared to the fourth-order results of 
Ref.~\cite{Epelbaum:2014efa}. This provides clear evidence of the corresponding 
two-pion exchange contributions with all low-energy constants being 
determined from pion-nucleon scattering.
The fifth-order corrections to nucleon-nucleon observables  
appear to be of a natural size which confirms the good convergence of the chiral expansion for nuclear forces. 
Furthermore, the obtained results provide strong support for the novel way of quantifying 
the theoretical uncertainty due to the truncation of the chiral expansion proposed in 
Ref.~\cite{Epelbaum:2014efa}. Our work opens up new perspectives for precision {\em ab initio} calculations 
in few- and many-nucleon systems and is especially relevant for ongoing 
efforts towards a quantitative understanding the structure of the three-nucleon force
in the framework of chiral effective field theory.
\end{abstract}

\maketitle

Chiral effective field theory (EFT) provides a solid foundation for analyzing low-energy 
hadronic observables in harmony with the symmetries of quantum chromodynamics (QCD), 
the underlying theory of the strong interactions. It allows one to derive nuclear forces 
and currents in a systematically improvable way order by order in the chiral expansion, based on a
perturbative  expansion in powers of 
$Q \in ( p/\Lambda_b, \; M_\pi/\Lambda_b)$, where $p$ refers to the magnitude of three momenta of the external particles,
$M_\pi$ is the pion mass and $\Lambda_b$  is the breakdown scale of chiral EFT \cite{Weinberg:1990rz}. 
Being combined with modern few- and many-body methods, the resulting framework  
based on solving the nuclear $A$-body Schr\"odinger equation with interactions between nucleons tied to
QCD via its symmetries represents nowadays a commonly accepted 
approach to  \emph{ab initio} studies of
nuclear structure and reactions, see Refs.~\cite{Epelbaum:2008ga,Machleidt:2011zz} for review articles.

Chiral power counting suggests that nuclear forces are dominated by pairwise interactions 
between the nucleons \cite{Weinberg:1990rz}, a feature that was known for long but could only 
be explained with the advent of chiral EFT. Many-body forces are suppressed by powers 
of the expansion parameter $Q$. Specifically, the chiral expansion of nucleon-nucleon (NN), three-nucleon 
(3NF) and four-nucleon (4NF) forces starts at the orders 
$Q^0$ (LO), $Q^3$ (N$^2$LO) and $Q^4$ (N$^3$LO), respectively, while 
next-to-leading (NLO) corrections involve two-body operators only.  
While accurate NN potentials at N$^3$LO have been available for about a decade \cite{Entem:2003ft,Epelbaum:2004fk}, 
the 3NF still represents one of 
the major challenges in the physics of nuclei and nuclear matter \cite{Hammer:2012id}. 
In particular, numerically exact calculations in the three-nucleon (3N) continuum, the most natural place to test the 3NF, 
have revealed that the spin-structure of 
the 3NF is not properly reproduced by the available models \cite{KalantarNayestanaki:2011wz}. Specifically, one observes 
clear discrepancies between theory and experimental data for various spin observables in nucleon-deuteron (Nd) scattering 
starting at $E_{N} \sim 50$ MeV which tend to increase with energy. In addition, there are a few discrepancies at 
low energies such as e.g.~the so-called $A_y$-puzzle, see \cite{KalantarNayestanaki:2011wz} for more details. 

In the framework of chiral EFT, the impact of the leading 3NF at N$^2$LO on three- and four-nucleon scattering, 
nuclear structure and reactions as well as  nuclear matter has been extensively 
studied  using different many-body techniques. In particular, the N$^2$LO 3NF was found to reduce the discrepancy 
for  $A_y$ in proton-$^3$He elastic scattering \cite{Viviani:2010mf}, to play a crucial role in 
understanding neutron-rich systems \cite{Wienholtz:2013nya} and the 
properties of neutron and nuclear matter, see \cite{Hammer:2012id} and references therein. 
Lattice simulations of light nuclei within the framework of chiral EFT also confirm 
the important role of the N$^2$LO 3NF \cite{Epelbaum:2009pd,Epelbaum:2011md,Epelbaum:2012qn}.
On the other hand, the $A_y$ puzzle in elastic Nd scattering is not resolved 
at N$^2$LO \cite{Viviani:2010mf}, and the existing discrepancies for spin observables in the 
3N continuum at medium and higher energies are beyond the expected theoretical 
accuracy at this order. It is, therefore, necessary to study corrections beyond the leading 3NF. The N$^3$LO 
contributions to the 3NF have been worked out recently and appear to be parameter-free \cite{Bernard:2007sp,Bernard:2011zr}. 
It was found, however, that the chiral expansion of the long- and intermediate-range parts of the 
3NF is not converged at this order due to large fifth-order (N$^4$LO) corrections associated with 
intermediate $\Delta$(1232) excitations \cite{Krebs:2012yv,Krebs:2013kha,Epelbaum:2014sea}. 
A resolution of the long-standing discrepancies in the 3N continuum 
will, therefore, likely require the knowledge of the nuclear Hamiltonian at N$^4$LO. 

In this Letter, we make an important step along this line and present the NN potential  
at fifth order in the chiral expansion based on the improved regularization framework introduced 
in Ref.~\cite{Epelbaum:2014efa}. In addition to constructing a new state-of-the-art chiral NN 
potential which leads to an excellent description of the data and is expected to provide a solid basis for 
future few- and many-body calculations, our study represents a highly nontrivial test of 
the convergence of the chiral expansion and of the new approach for estimating the theoretical 
uncertainty, a necessary ingredient of any EFT calculation \cite{Furnstahl:2014xsa}.  

\begin{figure}[tb]
\includegraphics[width=0.49\textwidth,keepaspectratio,angle=0,clip]{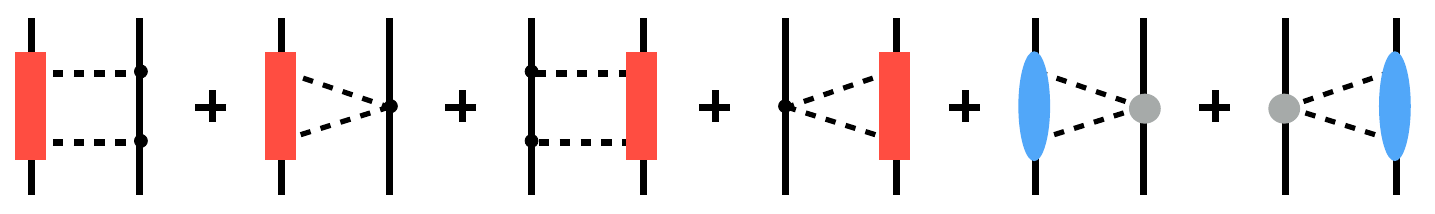}
    \caption{Fifth order contributions to the TPEP. Solid and dashed lines refer to nucleons and pions, respectively. 
Solid dots denote vertices from the lowest-order $\pi N$ effective Lagrangian. Filled (color online: red) rectangles,  
(color online: blue) ovals and grey circles 
denote the order $Q^4$, order $Q^3$ and order $Q^2$ contributions to $\pi N$ scattering, respectively.   
\label{Fig:diagrams} 
 }
\end{figure}

We first discuss the isospin-conserving part of the potential.  
As described in detail in Ref.~\cite{Epelbaum:2014efa}, the NN potential at N$^3$LO involves contributions 
from up to three-pion exchange  and contact interactions acting in S-, P- and D-waves and 
the mixing angles $\epsilon_1$ and $\epsilon_2$. When expressed in terms of physical values of the 
pion masses and pion-nucleon ($\pi N$) coupling constant, the expression for the one-pion exchange potential (OPEP) 
remains unchanged at N$^4$LO. On the other hand, the static two-pion exchange potential (TPEP) receives 
corrections at fifth order which are visualized in Fig.~\ref{Fig:diagrams}. 
The corresponding diagrams up to the two-loop level have been calculated recently using 
the Cutkosky rules \cite{Entem:2014msa}. We have independently calculated 
these contributions and have verified the expressions presented in that work. Next, 
one also needs to account for the Goldberger-Treiman discrepancy and the leading relativistic 
corrections to the order $Q^3$ TPEP. Notice that the latter were already taken into account in 
 Refs.~\cite{Epelbaum:2014efa}. Furthermore, in addition to the TPEP, one encounters 
subleading three-pion exchange diagrams at N$^4$LO. Similar to Refs.~\cite{Epelbaum:2004fk,Entem:2003ft,Epelbaum:2014efa}, 
we do not include the  three-pion exchange potential explicitly assuming that its effects can be 
well reproduced by contact interactions. This needs to be explicitly verified in future studies. 
A remarkable feature of the N$^4$LO NN potential is the absence of new isospin-conserving 
contact interactions. This can be traced back to parity conservation and to the fact that the N$^4$LO corresponds 
to an odd power of the expansion parameter, namely $Q^5$.  This feature allows one to unambiguously  probe 
the impact of the fifth-order TPEP in NN scattering. 

Our treatment of isospin-breaking (IB) corrections is limited to the one 
employed in the Nijmegen partial wave analysis (NPWA) \cite{Stoks:1993tb} which is used as input in our calculations, 
see Refs.~\cite{Epelbaum:2014efa} for more details. In particular, we do not include IB 
TPEP as it would affect the splittings between the isospin-$1$ neutron-proton (np) and proton-proton (pp) partial waves 
which, except for the $^1$S$_0$ channel, are \emph{not} independently determined from the data in Ref.~\cite{Stoks:1993tb}. 
The only new IB contribution we include compared to the N$^3$LO analysis 
of Refs.~\cite{Epelbaum:2014efa} is the momentum-dependent contact interaction in the $^1$S$_0$ channel, which 
results in $C_{1S0}^{\rm pp} \neq C_{1S0}^{\rm np}$ using the notation of that work.  

It remains to specify the values of the various parameters entering the potential. The $\pi N$ scattering 
amplitude at order $Q^4$ depends on certain combinations of low-energy constants (LECs) 
$c_i$, $\bar d_i$ and $\bar e_i$, see
\cite{Krebs:2012yv,Fettes:2000gb} for notations,
which can be determined from $\pi N$ scattering. 
Notice that at N$^3$LO, the TPEP only depends on the LECs $c_i$ and $\bar d_i$. In our work \cite{Epelbaum:2014efa}, 
we employed the empirical values of the LECs  $c_i$ and $\bar d_i$ as found in $Q^3$ analyses of 
$\pi N$ scattering, see table \ref{tab:LEC}. 
\begin{table}[t]
\caption{
Values of the LECs $c_i$, $\bar d_i$ and $\bar e_i$ in units of GeV$^{-1}$, GeV$^{-2}$ and GeV$^{-3}$, respectively, 
employed in the N$^3$LO NN potential of Ref.~\cite{Epelbaum:2014efa} and in the N$^4$LO NN potential of this work.
\label{tab:LEC}}
\smallskip
\begin{ruledtabular}
\begin{tabular}{@{\extracolsep{\fill}}rrr}
\noalign{\smallskip}
 LEC &  values used in \cite{Epelbaum:2014efa}   &  this work    
\smallskip
 \\
\hline 
$c_1$ & $-0.81$ & $-0.75$ \\ 
$c_2$ & $3.28$ & $3.49$ \\ 
$c_3$ & $-4.69$ & $-4.77$ \\ 
$c_4$ & $3.40$ & $3.34$ \\ 
$\bar d_1 + \bar d_2$ & $3.06$ & $6.21$ \\ 
$\bar d_3$ & $-3.27$ & $-6.83$ \\ 
$\bar d_5$ & $0.45$ & $0.78$ \\ 
$\bar d_{14} - \bar d_{15}$ & $-5.65$ & $-12.02$ \\ 
$\bar e_{14}$ & --- & $1.52$ \\ 
$\bar e_{17}$ & --- & $-0.37$ \\ 
\end{tabular}
\end{ruledtabular}
\end{table}
Specifically, the values in the second column of this table for $c_{1,3,4}$ refer to the central values 
determined from $\pi N$ scattering inside the Mandelstam triangle \cite{Buettiker:1999ap}, while the ones for 
$c_2$ and $\bar d_i$ are taken from the order $Q^3$ analysis of $\pi N$ phase shifts of Ref.~\cite{Fettes:1998ud} (fit 1).  
In this Letter, we adopt the values of the LECs listed in the last column of table \ref{tab:LEC}
which are taken from the order $Q^4$ fit of Ref.~\cite{Krebs:2012yv} based on the 
Karlsruhe-Helsinki partial wave analysis (PWA) of $\pi N$ scattering \cite{Koch:1985bn}. 
Notice that using the GWU PWA of Ref.~\cite{Arndt:2006bf}  or $\pi N$ scattering data as the input in the fits leads to 
slightly different values of the LECs \cite{Krebs:2012yv,Wendt:2014lja}. 

Here and in what follows, we adopt the same values for the pion and nucleon masses, 
pion decay constant, nucleon axial coupling constant and pion-nucleon coupling $g_{\pi N}$ as 
used in Refs.~\cite{Epelbaum:2014efa}. We also employ the same regularization framework. In particular, the 
OPEP and TPEP are regularized in $r$-space by multiplying with the function 
\beq
\label{NewReg}
f \left( \frac{r}{R} \right) = \left[1- \exp\left( - \frac{r^2}{R^2} \right)\right]^6\,,
\eeq
with the cutoff $R$ being chosen in the range of $R=0.8 \ldots 1.2\;$fm.  For contact interactions, 
we use a nonlocal Gaussian regulator in momentum space with the cutoff $\Lambda = 2 R^{-1}$, 
see \cite{Epelbaum:2014efa} for more details. We also adopt the same treatment of electromagnetic effects 
and relativistic corrections and
employ the same fitting strategy to determine the values of the LECs accompanying 
contact interactions as done in \cite{Epelbaum:2014efa}. In particular, we use np and pp 
phase shifts and mixing angles of the NPWA as input in our fits and define their error via 
\begin{eqnarray}
\Delta_X &=& \max \Big( \Delta^{\rm NPWA}_X, \; | \delta_X^{\rm NijmI} -
  \delta_X^{\rm NPWA} |,\\
&& {} \;\;\;\;\;\;\;\; \;\; | \delta_X^{\rm NijmII} -
  \delta_X^{\rm NPWA} |, \; | \delta_X^{\rm Reid93} -
  \delta_X^{\rm NPWA} | \Big)\,,
\nonumber
\end{eqnarray}
where $\delta_X$ denotes a given phase shift (or mixing angle) in the channel $X$,
$\Delta^{\rm NPWA}_X$ is the corresponding statistical error of the NPWA \cite{Stoks:1993tb},  while 
$\delta_X^{\rm
  NijmI}$,  $\delta_X^{\rm NijmI}$ and  $\delta_X^{\rm Reid93}$ denote
the results based on the Nijmegen I, II and Reid93 NN potentials of
Ref.~\cite{Stoks:1994wp} which can be regarded as alternative PWA. 
While $\chi^2/{\rm datum}$ for the description of the Nijmegen phase shifts 
calculated using the errors $\Delta_X$ defined above does, clearly, \emph{not} allow for 
statistical interpretation, see Ref.~\cite{Epelbaum:2014efa} for more details, it provides a useful 
tool to quantify the accuracy of the fits.

For all considered  values of the cutoff, namely $R=0.8$, $0.9$, $1.0$, $1.1$ and $1.2\;$fm, 
the resulting LECs are found to be natural  
and comparable in size with their N$^3$LO values given in Ref.~\cite{Epelbaum:2014efa}. 
We found that the inclusion of the fifth-order TPEP leads to a substantial improvement 
in the description of np and pp phase shifts (for hard cutoff choices). As an example, 
we show in table \ref{tab_chi2_2} the resulting $\chi^2/{\rm datum}$ for the description 
of the Nijmegen np and pp phase shifts using the cutoff $R=0.9$ fm, which was found 
 in Ref.~\cite{Epelbaum:2014efa} to yield most accurate results for NN
 observables. 
\begin{table}[t]
\caption{$\chi^2/{\rm datum}$ for the description of the Nijmegen
np and pp phase shifts \cite{Stoks:1993tb} at different
orders in the chiral expansion for the cutoff $R=0.9$ fm. Only those
channels are included which have been used in the N$^3$LO/N$^4$LO fits,
namely the S-, P- and D-waves and the mixing angles $\epsilon_1$ and
$\epsilon_2$.
\label{tab_chi2_2}}
\smallskip
\begin{ruledtabular}
\begin{tabular}{@{\extracolsep{\fill}}cccccc}
\noalign{\smallskip}
 $E_{\rm lab}$ bin &  LO   &  NLO   &  N$^2$LO   &  N$^3$LO   &  N$^4$LO  
\smallskip
 \\
\hline 
\multicolumn{5}{l}{neutron-proton phase shifts} \\ 
0--100 & 360 & 31 & 4.5 & 0.7  & 0.3\\ 
0--200 & 480 & 63 & 21 & 0.7  & 0.3\\ [4pt]
\hline 
\multicolumn{5}{l}{proton-proton phase shifts} \\ 
0--100 & 5750 & 102 & 15 & 0.8 & 0.3 \\ 
0--200 & 9150 & 560 & 130 & 0.7 & 0.6
\end{tabular}
\end{ruledtabular}
\end{table}
Notice that the additional IB  N$^4$LO contact term
 affects only np results. Switching it off leads to $\chi^2/{\rm
   datum} =0.5$  for the description of the np phase shifts in both
 energy bins.
Further, the residual cutoff dependence of the phase shifts appears, as expected, 
to be very similar at N$^4$LO and N$^3$LO. 
Also the error plots at N$^4$LO reveal a similar behavior to those at N$^3$LO
shown in Fig.~5 of that work, so that the estimation of the breakdown scale of $\Lambda_b = 600$ MeV
for $R=0.8 \ldots 1.0$ fm made in the N$^3$LO analysis of Ref.~\cite{Epelbaum:2014efa} remains valid at N$^4$LO. 

For the deuteron properties, the N$^4$LO predictions are very close to those at N$^3$LO (except for $P_D$ which is not observable), 
see table~\ref{tab_deut}, indicating a good convergence of the chiral expansion.
\begin{table}[t]
\caption{Deuteron binding energy $B_d$ (in MeV), asymptotic $S$ state
  normalization $A_S$ (in fm$^{-1/2}$) , asymptotic $D/S$ state ratio $\eta$, radius
  $r_d$ (in fm) and quadrupole moment $Q$  (in fm$^2$) at various orders in the chiral 
  expansion based on the cutoff $R=0.9$ fm in
  comparison with empirical information. Also shown is the $D$-state
  probability $P_D$ (in $\%$). Notice that $r_d$ and $Q$ are calculated
  without taking into account meson-exchange current contributions and
  relativistic corrections. The star indicates an input quantity.
  References to experimental data can be found in Ref.~\cite{Epelbaum:2014efa}.    
\label{tab_deut}}
\smallskip
\begin{ruledtabular}
\begin{tabular}{@{\extracolsep{\fill}}lllllll}
\noalign{\smallskip}
   &  LO & NLO  & N$^2$LO  & N$^3$LO  & N$^4$LO  & Empirical
\smallskip
 \\
\hline 
&&&&&&\\[-8pt]
$B_d$ & 2.0235 & 2.1987 & 2.2311 & 2.2246$^\star$ & 2.2246$^\star$ & 2.224575(9) \\ 
$A_S$ & 0.8333  & 0.8772  & 0.8865  & 0.8845  & 0.8844  & 0.8846(9) \\ 
$\eta$ & 0.0212 & 0.0256 & 0.0256 & 0.0255 &  0.0255 &  0.0256(4)\\ 
$r_d$& 1.990 &  1.968 & 1.966 & 1.972 &  1.972 &  1.97535(85)\\ 
$Q$ & 0.230 & 0.273 & 0.270 & 0.271 &   0.271 &  0.2859(3)\\ 
$P_D$ & 2.54 & 4.73 & 4.50 & 4.19 &  4.29 &  
\end{tabular}
\end{ruledtabular}
\end{table}
This feature holds true for all choices of the cutoff $R$. For $r_d$ and $Q$, the N$^4$LO predictions are 
in the range of $r_d=1.970 \ldots 1.981\;$fm and $Q=0.270\ldots 0.281\;$fm$^2$ for the cutoff 
variation of $R=0.8\ldots 1.2\;$fm. Taking into account the estimated size of the relativistic corrections and 
long-range meson-exchange current contributions, the observed spread in the values of $r_d$ and $Q$ is consistent 
with the estimated size of the corresponding short-range NN currents, see Ref.~\cite{Epelbaum:2014efa} and references therein. 

\begin{figure}[tb]
\includegraphics[width=0.49\textwidth,keepaspectratio,angle=0,clip]{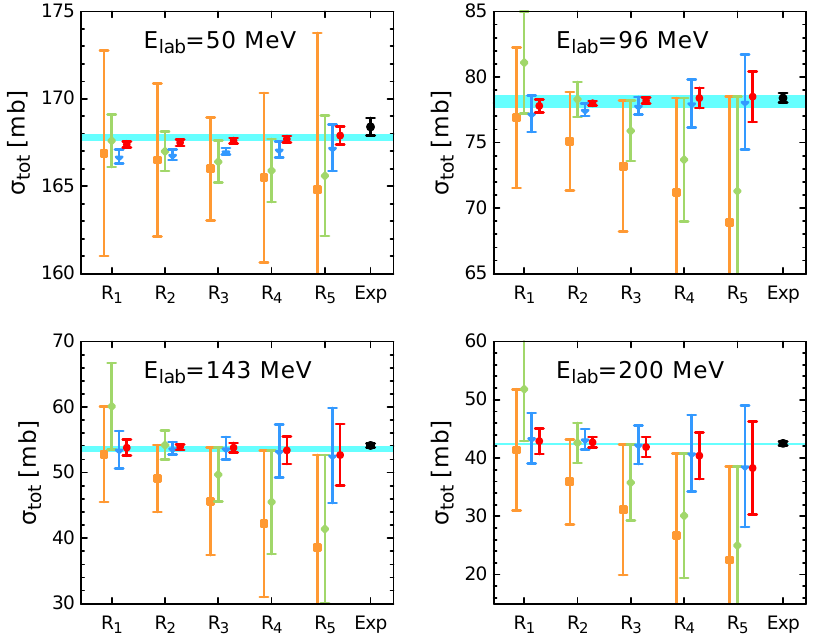}
    \caption{
Predictions for the np total cross section based on the improved chiral NN potentials at NLO 
(filled squares, color online: orange), N$^2$LO (solid diamonds, color online: green), N$^3$LO (filled triangles, 
color online: blue) and N$^4$LO (filled circles, 
color online: red) at the laboratory energies of $50$, $96$, $143$ and $200\;$MeV for
 the different choices of the cutoff: $R_1 = 0.8\;$fm, $R_2 = 0.9\;$fm, $R_3 = 1.0\;$fm, $R_4 = 1.1\;$fm and $R_5 = 1.2\;$fm. 
The horizontal band refers to the result of the NPWA with the uncertainty estimated as explained in the text. Also shown are experimental data of Ref.~\cite{Abfalterer:2001gw}.
\label{fig:sigtot} 
 }
\end{figure}

We now address the question of the theoretical uncertainty of our calculations due to the truncation 
of the chiral expansion. To this aim, we employ the approach proposed in Ref.~\cite{Epelbaum:2014efa} 
which is based on estimating the size of neglected higher-order contributions and does not 
rely on a cutoff variation. Specifically, the uncertainty $\Delta X^{\rm N^4LO} (p)$ of a N$^4$LO 
prediction $X^{\rm N^4LO}(p)$ for an observable $X(p)$, with $p$ referring to the center of mass momentum,  
is estimated via
\begin{eqnarray}
\label{def_error}
\Delta X^{\rm N^4LO} (p) &=& \max  \bigg( Q^6 \times \Big| X^{\rm
    LO}(p) \Big|, \\
&& {}  \;\;\;\;\;\; \; \; \; \;  Q^4 \times \Big|
  X^{\rm LO}(p) -   X^{\rm NLO}(p) \Big|, \nn  
&&{}  \;\;\;\;\;\; \; \; \; \;  Q^3 \times \Big|
  X^{\rm NLO}(p) -   X^{\rm N^2LO}(p) \Big|, \nn
&& {}  \;\;\;\;\;\; \; \; \; \;  Q^2 \times \Big|
  X^{\rm N^2LO}(p) -   X^{\rm N^3LO}(p) \Big|  , \nn
&& {}  \;\;\;\;\;\; \; \; \; \;  Q \times \Big|
  X^{\rm N^3LO}(p) -   X^{\rm N^4LO}(p) \Big|  \bigg)\,.
\nonumber
\end{eqnarray}
Here, $Q$ is the expansion parameter given by
\begin{equation}  
\label{expansion}
Q =\max \left( \frac{p}{\Lambda_b}, \; \frac{M_\pi}{\Lambda_b} \right)\,. 
\end{equation}
For the breakdown scale, we use the same values as in Ref.~\cite{Epelbaum:2014efa}, namely  
$\Lambda_b=600\;$MeV, $500\;$MeV and $400\;$MeV  
for $R=0.8\ldots 1.0\;$fm, $R=1.1\;$fm and $R=1.2\;$fm, respectively. The theoretical uncertainty at 
lower orders is estimated in a similar way as described in detail in \cite{Epelbaum:2014efa}. 
Fig.~\ref{fig:sigtot}  shows the resulting predictions for the np total cross section at different energies and 
for all cutoff choices. 
First, we observe that the predictions based on different values of the cutoff $R$ are consistent with 
each other with results corresponding to larger values of $R$ being less accurate due to a larger amount of 
cutoff artefacts. Secondly, our N$^4$LO predictions provide strong support for the new approach of error estimation. 
In particular, the actual size of the N$^4$LO corrections is in a good agreement with the estimated uncertainty 
at N$^3$LO \cite{Epelbaum:2014efa}. The somewhat larger N$^4$LO contributions at the lowest 
energy is to be expected and can be traced back to the adopted fitting strategy in the $^1$S$_0$ channel, 
see Ref.~\cite{Epelbaum:2014efa} for more details. Finally, our N$^4$LO results are 
in a very good agreement both with the NPWA and with the experimental data. 

\begin{figure}[tb]
\includegraphics[width=0.49\textwidth,keepaspectratio,angle=0,clip]{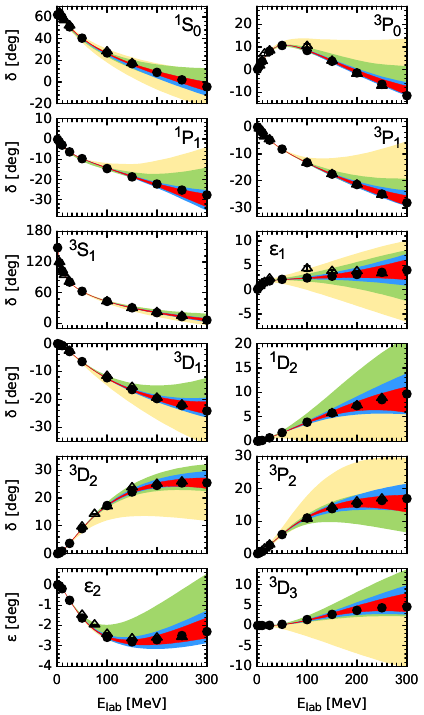}
    \caption{Results for the np S-, P- and D- waves and the mixing angles $\epsilon_1$,  $\epsilon_2$ up to N$^4$LO 
based on the
      cutoff of $R=0.9\,$fm in comparison with the NPWA \cite{Stoks:1993tb}  (solid dots) and the GWU 
      single-energy PWA \cite{Arndt:1994br} (open
      triangles).  The bands of increasing width show estimated theoretical uncertainty at 
      N$^4$LO (color online: red), N$^3$LO (color online: blue), N$^2$LO (color online: green) and NLO (color online: yellow).   
\label{fig:phases_conv} 
 }
\end{figure}

\begin{figure}[tb]
\includegraphics[width=0.49\textwidth,keepaspectratio,angle=0,clip]{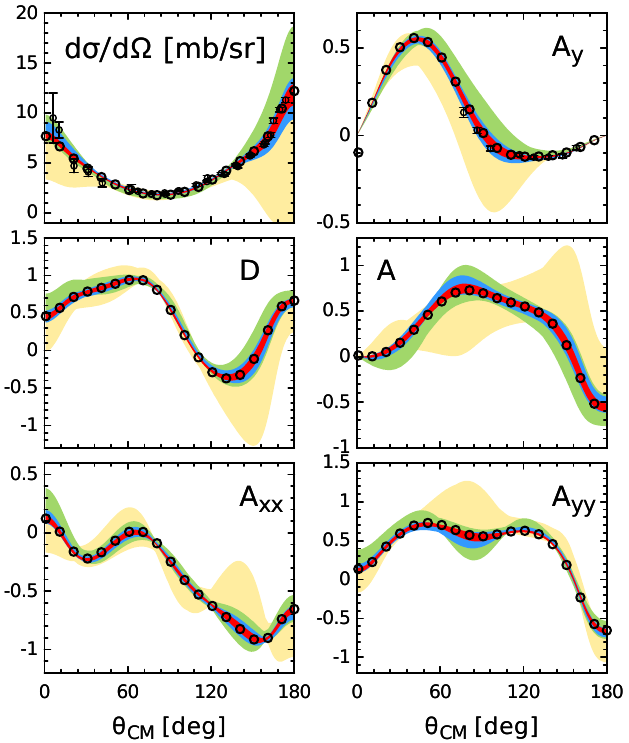}
    \caption{Predictions for the np differential
     cross section $d \sigma/ d \Omega$,  the vector analyzing power $A$,
     the polarization transfer coefficients $D$ and $A$ and the spin
     correlation parameters $A_{xx}$ and $A_{yy}$ at  $E_{\rm   lab} = 200\,$MeV calculated  up to N$^4$LO based on the
      cutoff of $R=0.9\,$fm.  Open circles refer to the result of
                  the NPWA \cite{Stoks:1993tb}. The bands of increasing width show estimated theoretical uncertainty at 
      N$^4$LO (color online: red), N$^3$LO (color online: blue), N$^2$LO (color online: green) and NLO (color online: yellow).   
For references to data see \cite{Epelbaum:2014efa}.
\label{fig:obs200} 
 }
\end{figure}

The above error analysis can be carried out for any observable of interest. Fig.~\ref{fig:phases_conv}
shows the estimated
uncertainty of the S-, P- and D-wave phase shifts and the mixing
angles $\epsilon_1$ and $\epsilon_2$ at NLO and higher orders in the chiral expansion based
on $R=0.9\,$fm. The various bands result by adding/subtracting the estimated
theoretical uncertainty, $\pm \Delta \delta (E_{\rm lab})$ and $\pm \Delta \epsilon (E_{\rm lab})$, to/from the
calculated results. Similarly, we show in Fig.~\ref{fig:obs200} our predictions for the various 
NN scattering observables at $E_{\rm lab}= 200\;$MeV. 
In all cases, we observe excellent agreement with the PWA and the available experimental data and 
confirm a good convergence of the chiral expansion. Furthermore, the N$^4$LO uncertainty bands lie within 
the N$^3$LO ones and describe the data. This provides a strong support for reliability of the proposed 
approach of error estimation. Similar conclusions follow from the results based on different values of the 
cutoff $R$ which are, however, less stringent due to lower accuracy of such calculations.

This work was supported by the EU (HadronPhysics3,
Grant Agreement n. 283286) under the Seventh Framework Programme of EU,
 the ERC project 259218 NUCLEAREFT and by the DFG and NSFC (CRC 110).

\end{document}